\documentclass{IEEEtran}

\usepackage{amsmath,amssymb,amsfonts}
\usepackage{algorithmic}
\usepackage{graphicx}
\usepackage{textcomp}
\usepackage{xcolor}
\def\BibTeX{{\rm B\kern-.05em{\sc i\kern-.025em b}\kern-.08em
    T\kern-.1667em\lower.7ex\hbox{E}\kern-.125emX}}

\usepackage{multicol}
\usepackage{setspace}
\usepackage{xspace}
\newcommand{\thesystem}{{\it MAD}\xspace}
\newcommand{\thesystembold}{{\textbf{\textit{MAD}}}\xspace}

\usepackage{hyperref}
\usepackage[hang,small,bf,tight]{subfigure} 
\usepackage{listings,lipsum}

\begin{document}

\title{Automatic Design-Time Detection of Anomalies in Migrating Monolithic Applications to Microservices}

\author{Valentim Rom\~{a}o, Rafael Soares, Luís Rodrigues, Vasco Manquinho\\INESC-ID\\Instituto Superior Técnico\\Universidade de Lisboa}

\maketitle

\begin{abstract}
  The advent of microservices has led multiple companies to migrate their monolithic systems to this new architecture. When decomposing a monolith, a functionality previously implemented as a transaction may need to be implemented as a set of independent sub-transactions, possibly executed by multiple microservices. The concurrent execution of decomposed functionalities may interleave in ways that were impossible in the monolith, paving the way for anomalies to emerge. The anomalies that may occur critically depend on how the monolith is decomposed. The ability to assess, at design time, the anomalies that different decompositions may generate is key to guide the programmers in finding the most appropriate decomposition that matches their goals. This paper introduces \thesystem, the first framework for automatically detecting anomalies that are introduced by a given decomposition of a monolith into microservices. \thesystem operates by encoding non-serializable executions of the original functionalities as an SMT formula and then using a solver to find satisfiable assignments that capture the anomalous interleavings made possible by that specific decomposition. We have applied \thesystem to different benchmarks and show that it can identify precisely the causes of potential anomalous behavior for different decompositions. 
\end{abstract}

\section{Introduction}

Microservices have emerged as a promising architecture for implementing large-scale applications. When using microservices, applications are designed as a set of loosely coupled components that may be easily developed and maintained by independent teams~\cite{MicroPatternsBook, microservices_journey}. The adoption of this architectural style has led many companies, including large companies such as Amazon, Netflix, and Uber, to migrate applications that have been previously implemented as monoliths to microservices~\cite{microservice-examples,study_microservices, microservices_netflix, microservices_linkedin, microservices_soundcloud, Microservices_migration1, Microservices_migration2}.

Unfortunately, migrating an application to the microservice architecture is not a trivial task~\cite{challenges_migrating_to_microservices, microservices_implications}. Functionalities that have been designed in the monolith to execute as a single ACID transaction may be required to execute as a sequence of independent transactions after the migration, each implemented by a different microservice. This breaks the isolation among functionalities that are executed concurrently, leading to anomalous application behavior. Handling anomalous behavior is costly as it often requires the implementation of additional code, such as compensating actions~\cite{Sagas}, to correct the undesired effects of the loss of isolation and atomicity. In some cases, these costs outweigh the advantages of microservices, forcing developers to revert the application to a monolith~\cite{fromMicro2Mono}.

The amount of anomalies a migration may foster is heavily dependent on how the monolith is decomposed, namely, on how the domain entities are assigned to the different microservices. Therefore, before deciding on a given decomposition, it is critical to understand how many anomalies a decomposition may generate, as dealing with anomalies typically dominates the migration cost~\cite{ComplexityMetric}. The importance of estimating the complexity associated with a monolith decomposition has been recognized in the literature~\cite{MetricsRefinement}. However, previous works in this direction do not identify concrete anomalies. The costs of avoiding or compensating anomalies varies greatly depending on the anomaly's type~\cite{ConsistencyRationing, ConsistencySLA}. This information not only helps developers to choose the most cost-efficient decomposition, but also raises awareness of the concrete challenges they will face during the decomposition process. 


In this paper, we propose the \textit{Microservices Anomaly Detector} (\thesystem), a new framework to analyze the anomalies that result from the decomposition of a monolith into microservices considering bounded executions of the system. \thesystem takes a monolithic application, a target decomposition, and the application's SQL schema to automatically generate, from the transaction that implements a functionality in the monolith, the set of independent transactions that execute the same functionality in the microservices. Then, \thesystem encodes the interleavings of these transactions that correspond to non-serializable executions of the original functionalities as a \textit{Satisfiability Module Theories} (SMT) formula. Using the Z3 solver~\cite{Z3_solver}, \thesystem finds the satisfiable assignments of this formula that capture the interleaving that may generate anomalous behaviors. To the best of our knowledge \thesystem is the first system that is able to precisely detect the anomalies that will result from the decomposition of monoliths in microservices.

Because \thesystem performs an exhaustive search on the space of all transaction interleavings (that respect a predefined bound 
, as discussed later), the time it takes to serially analyze a decomposition may be large. To circumvent this limitation, \thesystem implements a novel divide-and-conquer technique capable of parallelizing the search. This makes \thesystem suitable to be applied to non-trivial code bases. To illustrate the power of \thesystem, we have applied it to seven different benchmarks, and we show how it may guide developers to find suitable decompositions of the monolith. 

In summary, the contributions of this paper are as follows:

\begin{itemize}
\item We propose a technique to formulate the problem of finding anomalies in a microservice decomposition of a monolith as an SMT problem.
\item We propose a strategy to parallelize the task of finding the satisfiable assignments that capture anomalies.
\item We propose a technique to describe the satisfiable assignments to the developer in a meaningful manner, in particular, by classifying the type of anomaly that may occur and the entities and functionalities involved.
\item We present an experimental evaluation of the resulting system with 7 different benchmarks.
\end{itemize}

\section{Background}
\label{sec:related_work}

The problem of detection anomalies in transactional applications has been addressed in the literature using different approaches, including testing and validation. 

Relevant examples of testing tools are systems such as \textit{MonkeyDB}~\cite{MonkeyDB} and \textit{Cobra}~\cite{COBRA}, that detect anomalies by using a black box approach. These tools generate sets of test inputs and capture the application output, comparing it against the expected behavior of the application. An anomaly is detected when a mismatch is found between the obtained and the expected outputs. These approaches are generic because they do not require access to the source code. Unfortunately, there is no guarantee that all possible interleavings are tested, hence, some anomalies may pass unnoticed. More importantly, when applied to our goals, these tools require a target decomposition to be implemented before it can be tested. In opposition, we aim at detecting problematic decompositions at design time, such that programmers avoid implementing decompositions that generate an undesirable amount of anomalies.

Early validation work for transactional systems, such as~\cite{fekete2005,Sudhir2007}, assume a single database and a set of transactions that can be executed in any order. More recent works such as \textit{ANODE}~\cite{ANODE} and \textit{CLOTHO}~\cite{CLOTHO} consider distributed storage but assume that all storage nodes replicate all entities, while also assuming that transactions can execute independently of each other. When considering the decomposition of monolith into microservices, one must take into account that individual sub-transactions were originally sequential operations of a single functionality in the monolith: 
this imposes constrains on the order of execution of these sub-transactions and the values read and written by each sub-transaction as well, which need to be taken into account in the analysis. Also, these systems take as input a fixed set of transactions, and are not able to automatically derive the sub-transactions that result from the decomposition of a functionality.  \thesystem does not suffer from these limitations: i) it can model systems where each microservice has its own storage; ii) it takes the ordering constraints that result from the monolith decomposition into account, and; iii) it can automatically chop the functionalities from the monolith into multiple sub-transactions, for different combinations of domain entities into microservice aggregates.

Several works have addressed the problem of how to aggregate the domain entities when migrating from a monolith to microservices. Nunes et al.~\cite{Mono2Micro_rito} aggregate domain entities based on the transactional contexts of the monolithic application. Brito et al.~\cite{Mono2Micro_sac} aggregate domain entities using topic modeling.  Mono2Micro~\cite{Mono2Micro_ibm} and FoME~\cite{Mono2micro_microservices_benchmarks4}  use run-time traces to cluster the domain entities. None of these tools capture the anomalous behaviors that may result from concurrent executions of functionalities. \thesystem complements these works by identifying the anomalies that result when a given decomposition is used in the migration process.

\section{Example}
\label{sec:Example}

\begin{figure*}[t]
    \centering
    \subfigure[Original transactions]{\includegraphics[width=0.3\textwidth]{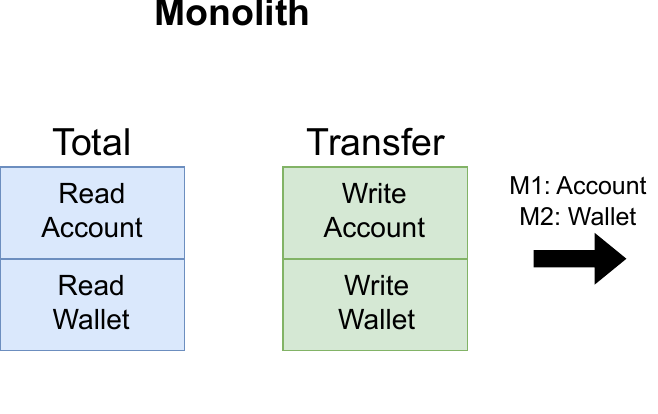}\label{fig:migration_example_a}}~\;~\;~\;
    \subfigure[Sub-transactions]{\includegraphics[width=0.3\textwidth]{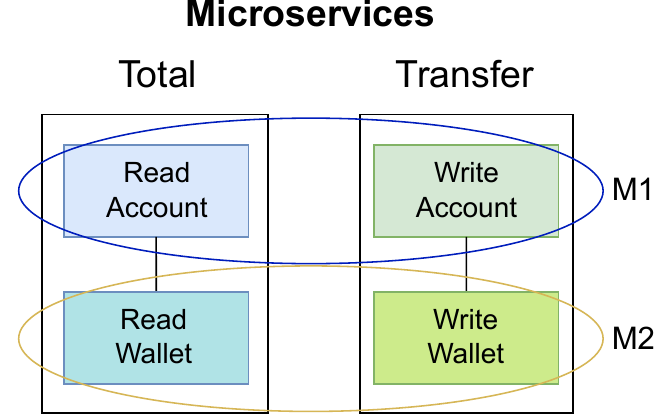}\label{fig:migration_example_b}}~\;~\;~\;
    \subfigure[Possible interleaving]{\hspace{1cm}\includegraphics[width=0.15\textwidth]{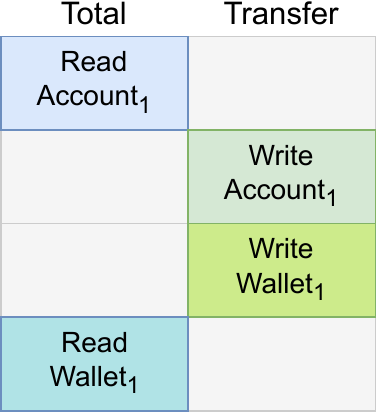} \hspace{1cm}\label{fig:anomaly_interleaving}}
    \caption{Example of how two functionalities can be divided when migrating from monolith to microservices and of an interleaving that leads to an anomaly.} \label{fig:simple-example}
\end{figure*}

Figure~\ref{fig:simple-example} illustrates an example of how a given microservices decomposition originates a previously unexisting  interleaving. 
In this scenario, there are two entities (\texttt{Account} and \texttt{Wallet}), two transactions (\textsf{Total} and \textsf{Transfer}), and a decomposition where \texttt{Account} is managed by microservice $M_1$ and \texttt{Wallet} is managed by microservice $M_2$. 
\textsf{Total} gets the total amount of a client's money (its account balance plus its wallet balance). \textsf{Transfer} withdraws an amount of funds from the client's account balance and deposits it in their wallet. Considering that a client's balance can only be transferred from their account to their wallet, it is expected that their total amount of funds of a client always remains the same. Since entities \texttt{Account} and \texttt{Wallet} are in different microservices, \textsf{Total} and \textsf{Transfer} would be split into sub-transactions, each of which executes in the microservice associated with the entity they are accessing. 

Transaction chopping allows for interleavings between sub-trans\-ac\-tions, which can lead to anomalies that were not possible in the monolithic version. One example is shown in Figure~\ref{fig:anomaly_interleaving}.
In this case, the execution of \textsf{Transfer} interleaves with the execution of \textsf{Total}. \textsf{Total} sees an older version of a client's account (\texttt{Account\textsubscript{1}}), implying that \textsf{Total} is executed before \textsf{Transfer}. However, \textsf{Total} sees the new version of the client's wallet (\texttt{Wallet\textsubscript{1}}), whose balance was already updated by \textsf{Tranfer}, implying that \textsf{Total} is executed after \textsf{Transfer}. There is no serial order of the two functionalities that may lead to this execution. From this execution, one could incorrectly observe that the total funds of the client have have changed, something impossible in this scenario.

\section{Microservices Anomaly Detector}

This section describes the \textit{Microservices Anomaly Detector} (\thesystem), a framework to automatically detect anomalies that would result from implementing a given microservices decomposition of a monolithic application. \thesystem works by comparing the feasible executions in the monolith and the feasible executions in a target microservice architecture, to detect anomalies under bounded executions that result from a specific decomposition while avoiding false positives.

\subsection{Overview}
\label{sec:mad_overview}

\thesystem builds an abstract representation of the decomposed monolith based on the original monolith code and a user-provided decomposition. Using this abstract representation, \thesystem builds SMT formulas that encapsulate the anomalous interleavings made possible due to the decomposition. Finally, the satisfiable assignments found by the SMT solver are classified based on their anomaly types and are returned to the developer, along with other metrics. In the following, we provide more detail about each of these steps.

\begin{figure*}[t]
     \centering
     \includegraphics[width=.9\textwidth]{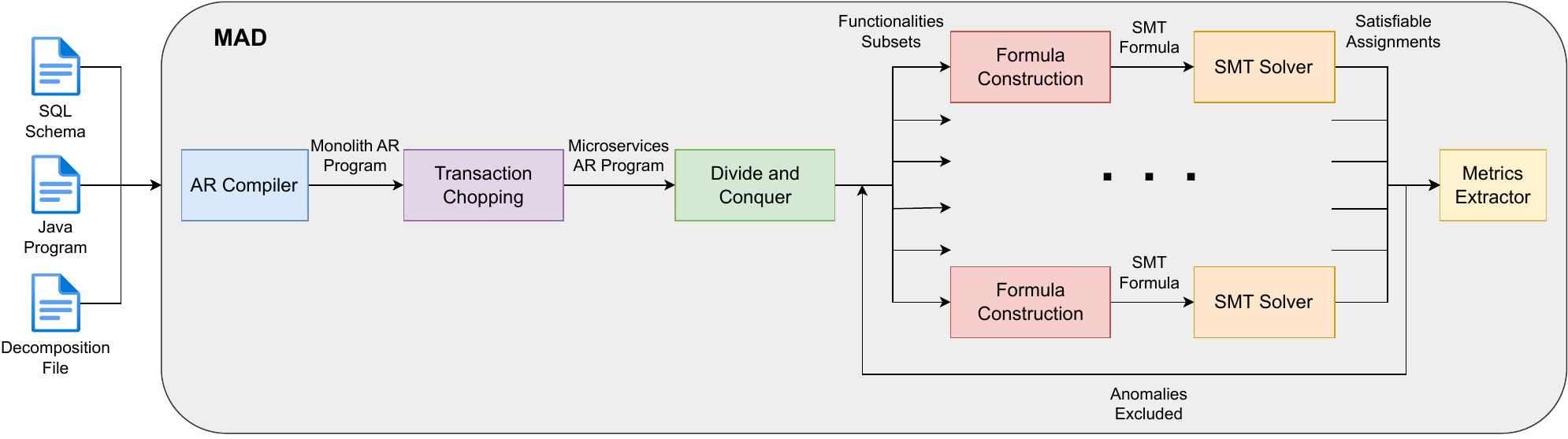}
     \caption{\thesystembold's pipeline.} \label{fig:MAD_pipeline}
\end{figure*}

\medskip\noindent \textit{Input: }\thesystem takes as input a monolithic implementation of an application (including its  SQL schema and source code) and a high-level description of how the monolith is decomposed into multiple microservices. In the current version, the source code must be a \textit{Java program} written using the JDBC syntax (i.e., that uses SQL queries to access the entities, which are maintained by the application in a database). As an illustration, the SQL schema and the Java code for the scenario from Figure~\ref{fig:simple-example} is depicted in Listings~\ref{lst:ExampleSQL} and~\ref{lst:ExampleJava}, respectively. Although the current prototype only supports Java, the framework has been designed such that it can be extended to support additional programming languages.  The decomposition of the monolith is expressed as the clustering of the domain entities into aggregates~\cite{Mono2Micro_rito} (entities grouped in the same cluster are assumed to be managed by the same microservice) and is represented by a JSON file (\textit{Decomposition File}). The JSON file used in the example scenario is depicted in Listing~\ref{lst:ExampleJSON}. Using this input, \thesystem executes the pipeline presented in Figure~\ref{fig:MAD_pipeline}, which is composed of the following sequence of steps:

\begin{figure}[t]
\begin{center}
\begin{spacing}{1}
\begin{lstlisting}[frame=lines,language=SQL,basicstyle=\scriptsize\ttfamily,caption={Example of \thesystem's input SQL schema file.},captionpos=b,label=lst:ExampleSQL]
CREATE TABLE Account (
  clientId INT,
  balance INT,
  PRIMARY KEY (clientId)
);

CREATE TABLE Wallet (
  clientId INT,
  balance INT,
  PRIMARY KEY (clientId)
);
\end{lstlisting}
\end{spacing}
\end{center}
\end{figure}

\begin{figure}[t]
\begin{center}
\begin{spacing}{1}
\begin{lstlisting}[frame=lines,language=Java,basicstyle=\tiny\ttfamily,caption={Example of \thesystem's input Java file.},captionpos=b,label=lst:ExampleJava]
public class exampleScenario {
  private Connection connect = null;
  private int _ISOLATION = Connection.TRANSACTION_READ_COMMITTED;
  private int id;
  Properties p;
  
  public exampleScenario(int id) {
    this.id = id;
    p = new Properties();
    p.setProperty("id", String.valueOf(this.id));
    Object o;
    try {
      o = Class.forName("MyDriver").newInstance();
      DriverManager.registerDriver((Driver) o);
      Driver driver = DriverManager.getDriver("jdbc:mydriver://");
      connect = driver.connect("", p);
    } catch (InstantiationException | IllegalAccessException | 
        ClassNotFoundException | SQLException e) {
      e.printStackTrace();
    }
  }
  
  public void Total(int clientId) throws SQLException {
    PreparedStatement stmt1 = connect.prepareStatement(
                            "SELECT balance FROM Account"+
                            " WHERE clientId = ?");
    stmt1.setInt(1, clientId);
    ResultSet rs = stmt1.executeQuery();
    rs.next();
    int account_balance = rs.getInt("balance");
  
    PreparedStatement stmt2 = connect.prepareStatement(
                            "SELECT balance FROM Wallet"+
                            " WHERE clientId = ?");
    stmt2.setInt(1, clientId);
    ResultSet rs2 = stmt2.executeQuery();
    rs2.next();
    int wallet_balance = rs2.getInt("balance");

    int total_money = account_balance + wallet_balance;
  }
  
  public void Transfer(int clientId, int accountBalance, 
                        int walletBalance, int amount) throws SQLException {
    PreparedStatement stmt1 = connect.prepareStatement(
                                "UPDATE Account SET balance = ?"+
                                " WHERE clientId = ?");
    stmt1.setInt(1, accountBalance - amount);
    stmt1.setInt(2, clientId);
    stmt1.executeUpdate();
  
    PreparedStatement stmt2 = connect.prepareStatement(
                                "UPDATE Wallet SET balance = ?"+
                                " WHERE clientId = ?");
    stmt2.setInt(1, walletBalance + amount);
    stmt2.setInt(2, clientId);
    stmt2.executeUpdate();
  }
}
\end{lstlisting}
\end{spacing}
\end{center}
\end{figure}

\begin{figure}[t]
\begin{center}
\begin{spacing}{1}
\begin{lstlisting}[frame=lines,basicstyle=\scriptsize\ttfamily,caption={JSON file with a microservices decomposition.},captionpos=b,label=lst:ExampleJSON]
{
  "M1": ["Account"],
  "M2": ["Wallet"]
}
\end{lstlisting}
\end{spacing}
\end{center}
\end{figure}

\medskip \noindent\textit{Step 1 (AR Compiler):} The source code is compiled to an \textit{abstract representation} (AR) that captures how the functionalities access the domain entities. In the AR, each functionality is represented by a sequence of read and/or write operations on domain entities (\textit{Monolith AR Program}). The AR compiler is the only component of \thesystem that needs to be extended if support for additional programming languages is required.

\medskip \noindent\textit{Step 2 (Transaction Chopping):} From the AR of the functionalities, an AR of the microservice decomposition is automatically generated (\textit{Microservices AR Program}) by chopping the functionalities code into multiple sub-sequences, where each sub-sequence only accesses domain entities from the same aggregate. Each sub-sequence captures a transaction that is executed in a single microservice. Note that, in the monolith, each functionality would be implemented using a single transaction. In the microservice decomposition, sub-sequences that are part of the same parent functionality are executed as a sequence of independent transactions.

\medskip \noindent\textit{Step 3 (Divide and Conquer):} Often, the AR program is too complex to be represented in a single encoding that can be effectively analyzed. Employing a divide and conquer strategy, \thesystem generates subsets of the functionalities (\textit{Functionalities Subsets}). Analyzing these subsets independently and in parallel allows the analysis of the whole problem to terminate in reasonable time.

\medskip \noindent\textit{Step 4 (Formula Construction):} From the AR and the subsets of functionalities, \thesystem generates SMT formulas encoding the program and whose satisfiable assignments are the possible interleavings between functionalities that can lead to anomalies in the decomposition. An SMT formula is built for each subset of functionalities, and their satisfiability can be checked by an SMT solver.

\medskip \noindent\textit{Step 5 (SMT Solver):} \thesystem uses Z3~\cite{Z3_solver} to check the satisfiability of the SMT formulas. Satisfiable assignments correspond to cyclic graphs, with the vertices being operations and the edges the relations between operations, representing unserializable executions of functionalities~\cite{dependency_graphs}. These cycles have their length bounded by a system parameter denoted the \emph{Maximum Cycle Length} (MCL), defined as the maximum number of edges to be considered when looking for satisfiable assignments. This parameter is required to prevent Z3 from continuously trying to find new satisfiable assignments which are supersets of previously found cyclic graphs.

\medskip \noindent\textit{Step 6 (Metrics Extractor):} After all satisfiable assignments are found for all SMT formulas, \thesystem processes the anomalies found and extracts complexity metrics to report to the user. These metrics include the total number of anomalies, dividing them in \textit{core anomalies} (represented by cyclic graphs with the minimum cycle length required to represent an anomaly) and their \textit{extensions} (represented by cyclic graphs, which are supersets of the \textit{core anomalies} graphs), the anomalies types, and the entities, functionalities, and sub-transactions involved.

\noindent Next, we describe each of these steps in more detail.

\subsection{Abstract Representation}

\begin{figure}[t]
     \centering
     \includegraphics[width=.47\textwidth]{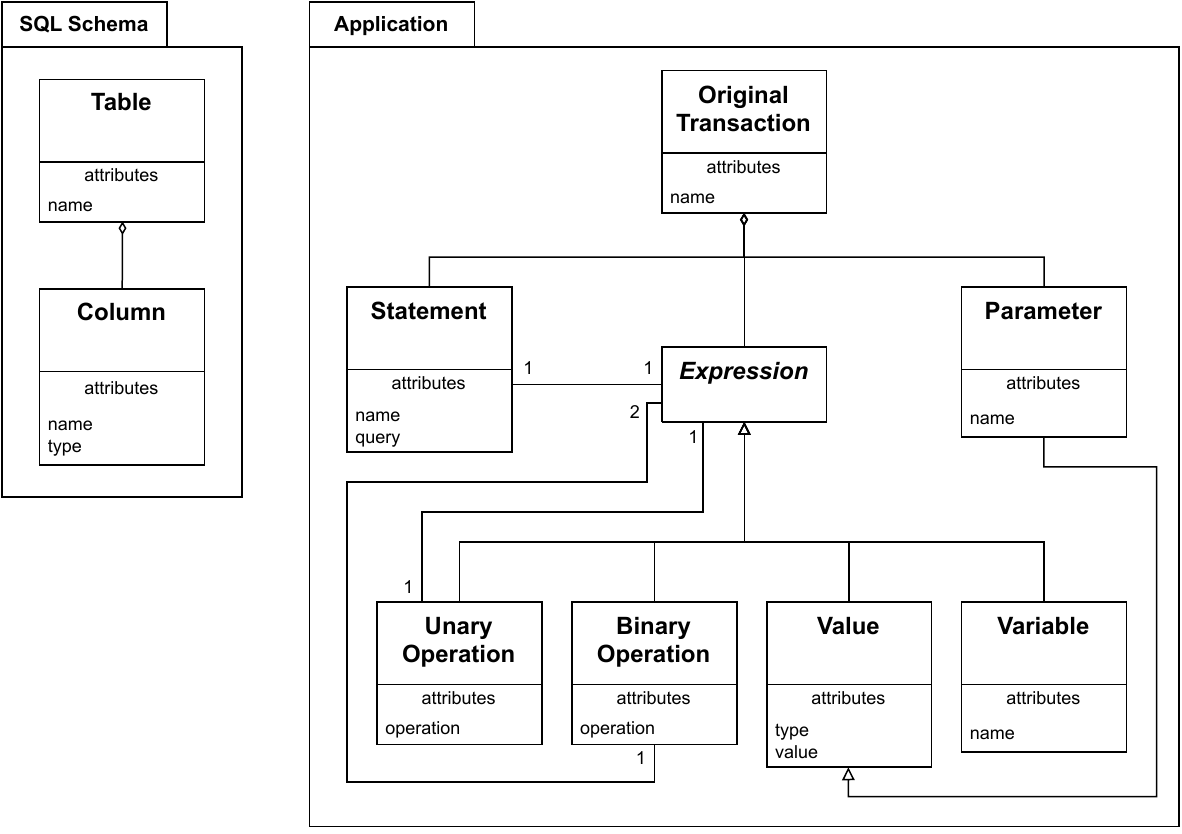}
     \caption{Monolith AR structure.} \label{fig:AR_structure}
\end{figure}

\thesystem receives as input the  Java source code of the monolith and compiles it to an abstract representation (AR), originating the \textit{Monolith AR Program}. The AR facilitates the extraction of information, including the transactions, types of parameters, and execution order. In Figure~\ref{fig:AR_structure}, we present the structure of the monolith AR.

To represent the SQL schema, \thesystem uses two elements, \textit{Table} and \textit{Column}. \textit{Table} is represented by a \textit{name} and a list of \textit{Columns}. \textit{Column} is represented by a \textit{name}, and a \textit{type} (int, real, string, boolean).

The application implementation is represented by a set of \textit{Original Transactions}. Each \textit{Original Transaction} captures the code of a functionality in the monolith and has a \textit{name}, a list of \textit{Statements}, a list of \textit{Expressions}, and a list of \textit{Parameters}. Each \textit{Statement} is represented by a \textit{name}, an SQL \textit{query} (select, update, insert or delete), and a \textit{path condition}, which is an expression that associates the execution of the statement with a condition in the program if it occurs inside a conditional block. The \textit{Expressions} can be of four types: \textit{Unary Operation}; \textit{Binary Operation}; \textit{Value}; and \textit{Variable}. The \textit{Unary} and \textit{Binary Operations} have one \textit{operation} and the \textit{Expression(s)} to which the \textit{operation} is applied, respectively. The \textit{Value} expressions represent the static values of the Java program using a \textit{type} and a \textit{value}. At last, the \textit{variable} expressions represent the variables used in the Java program using the variable \textit{name}. These variables include the ones used for the instructions, to store the rows read, and to hold the values of columns read from the database. To conclude, the \textit{Parameters} are a specific type of \textit{Value} expression, which are represented additionally by their \textit{name}.

\subsection{Microservices Decomposition AR}

From the Monolith AR Program and the JSON Decomposition File, \thesystem proceeds to create the \textit{Microservices AR Program}. This step consists of applying a transaction chopping algorithm to the original functionalities of the monolith to transform each of them in a sequence of sub-transactions. The \textit{sub-transactions} are represented in the AR by the following attributes: \textit{name}; list of SQL \textit{operations}; their \textit{original transaction name}; and their \textit{microservice name}.

The transaction chopping algorithm works as follows. For each original transaction, \thesystem iterates over the original sequence of operations and generates sub-transactions based on the entities that are accessed by these operations, assuming that each operation only accesses one entity. In each iteration, \thesystem evaluates if the currently accessed entity belongs to a different microservice than the previously accessed entity. If so, a new sub-transaction is created, starting with the current operation since it would execute in a different microservice. Otherwise, the current operation is added to the most recent sub-transaction created. After the algorithm finishes, the resulting Microservices AR Program is shown to the programmer. This transaction-chopping algorithm alleviates the programmers from having to divide the transactions by hand. Yet, the programmer may fine-tune the chopping, for instance, by re-ordering operations in the code of the original transaction (to automate such optimizations is outside the scope of this work).

As an example, consider the scenario from Figure~\ref{fig:simple-example}. By applying the previously described algorithm, from the original transactions depicted in Figure~\ref{fig:migration_example_a} we obtain a representation of the microservices version depicted in Figure~\ref{fig:migration_example_b}. Each sub-transaction only accesses one microservice and is executed as an independent transaction. However, it is still part of the original transaction execution flow.

\subsection{Divide and Conquer}

The first step of the \textit{Divide and Conquer} strategy is to generate all the combinations of functionalities (original transactions) of size smaller than the system parameter \emph{Maximum Cycle Length} (MCL). Note that, since an anomalous interleaving requires that at least two operations belong to the same instance of a functionality, at most $\textit{MCL}-1$ functionalities can be involved in a cycle without exceeding the MCL.

Let $n$ be the number of functionalities. The number of generated combinations is  given by $\sum_{x=1}^{\textit{MCL}-1}\ _{n}C_x$. \thesystem starts by creating one thread for each combination of size 1 (i.e., involving a single functionality). Each thread explores the possible interleavings between the operations of each functionality. When all these threads finish their analysis, the process is repeated for size~2 combinations, already excluding the anomalies involving only size~1 combinations. This process continues until all combinations of size $\textit{MCL}-1$ are analyzed.

Our approach has resemblances to the well-known Cube-and-Conquer algorithm~\cite{DBLP:conf/hvc/HeuleKWB11}, except that \thesystem splits the search space explicitly by taking advantage of domain knowledge. Hence, our approach is closer to a classic divide-and-conquer algorithm. Note that each thread analyzes a different combination of functionalities. Moreover, constraints are added such that solutions with $k-1$ functionalities are excluded when solving a formula involving $k$ functionalities. Although \thesystem generates one formula per combination, the search is done incrementally. Furthermore, the SMT formula handled in each thread is much smaller than solving the whole problem (i.e., considering all functionalities at once), allowing the parallelization of the search process and providing shorter solving times.

\subsection{SMT Encoding}

\thesystem encodes the AR program into an SMT formula such that any satisfiable assignment corresponds to an anomaly. We first provide a high level view of the SMT formulas and the information required to be represented in it, and later present the detailed representation.

We base our anomaly detection on Adya et al. definition of anomaly~\cite{dependency_graphs}. An execution is represented as a graph, where vertices correspond to operations belonging to transactions and edges represent data dependencies between operations. An anomaly is present if the graph is cyclic and contains at least two dependency edges and at least two operations belonging to the same transaction, also represented as an edge. As such, we must encode the transactions' operations and the data dependencies that different operations may present.

Different types of dependency edges may generate different types of anomalies. As such, we encode information regarding the type of dependencies between operations, which we use to identify which type of anomaly is represented in the cycle.

Furthermore, two operations might belong to different transactions, executing in different microservices, but belong to the same functionality. Operations in the same functionality must also execute in isolation to provide equivalent execution to the monolith, and interleavings may generate new anomalies. Therefore, we encode information regarding the functionality and microservice associated with each operation.

First, we represent the basic elements of the system (operations, sub-transactions, original transactions, and microservices). Second, the associations between the basic elements (e.g., to which sub-transaction does a given operation belong). Third, the consistency guarantees offered by the environment where the microservices execute. Fourth, the possible types of relations between operations. Lastly, the format of the cycles \thesystem wants the SMT solver to find.

\subsubsection{Representation of the Basic Elements}

For the representation of the basic elements, \thesystem uses sorts, which can be considered as types of objects. To represent instances of the operations, sub-transactions, and original transactions, \thesystem uses three sorts, \textit{O}, \textit{T}, and \textit{F}, respectively. 
Based on the AR of the program, \thesystem defines a unique name for each operation, sub-transaction, original transaction, and microservice. These unique identifiers are declared using the following sorts: \textit{ONames} for operations; \textit{TNames} for sub-transactions; \textit{FNames} for original transactions; and \textit{MNames} for microservices. Associated with the basic elements, there are functions and predicates used in the formulas. The most relevant ones are listed in Table~\ref{tab:functions} and Table~\ref{tab:predicates}.

\begin{table*}[t]
\centering
\caption{Key functions used in formulas.}\label{tab:functions}
\scalebox{.8}{
\begin{tabular}{|p{.2\textwidth}|p{.95\textwidth}|}
\hline
\textbf{Function} & \textbf{Description} \\
\hline\hline
\textbf{otime}$(O)$ & returns the instant of time when operation $O$ executes\\\hline
\textbf{tname}$(T)$ & returns the sub-transaction name (\textit{TName}) of sub-transaction $T$\\\hline
\textbf{oname}$(O)$ & returns the operation name (\textit{OName}) of operation $O$\\\hline
\textbf{fname}$(F)$ & returns the original transaction name (\textit{FName}) of original transaction $F$\\\hline
\textbf{mname}$(O)$ & returns the name of the microservice (\textit{MName}) where operation $O$ is executed\\\hline
\textbf{parent}$(O)$ & returns the instance of sub-transaction (\textit{T}) where operation $O$ belongs\\\hline
\textbf{origtx}$(O)$ & returns the instance of original transaction (\textit{F}) where operation $O$ belongs\\\hline
\end{tabular}
}
\end{table*}

\begin{table*}[t]
\centering
\caption{Key predicates used in formulas.}\label{tab:predicates}
\scalebox{.8}{
\begin{tabular}{|p{.15\textwidth}|p{\textwidth}|}
\hline
\textbf{Predicate} & \textbf{Description} \\
\hline\hline
\textbf{is\_update}$(O)$ & returns true if operation $O$ is an update operation (update, insert, delete) \\\hline
\textbf{ST}$(O,O)$ & returns true if both operation's instances $(O,O)$ belong to the same instance of sub-transaction (\textit{ST})\\\hline
\textbf{SOT}$(O,O)$ & returns true if both operation's instances $(O,O)$ belong to the same instance of original transaction (\textit{SOT})\\\hline
\textbf{WR}$(O,O)$ & returns true if there is a write followed by a read dependency (\textit{WR}) between operation's instances $(O,O)$\\\hline
\textbf{RW}$(O,O)$ & returns true if there is a read followed by a write dependency (\textit{RW}) between operation's instances $(O,O)$ \\\hline
\textbf{WW}$(O,O)$ & returns true if there is a write followed by a write dependency (\textit{WW}) between operation's instances $(O,O)$ \\\hline
\textbf{vis}$(O,O)$ & returns true if the visibility effects of the first instance are visible to the second instance of operation's instances $(O,O)$\\\hline
\textbf{ar}$(O,O)$ & returns true if the first instance is executed before the second instance of the operation's instances $(O,O)$\\\hline  
\textbf{D}$(O,O)$ & returns true if there is any dependency relation between operation's instances $(O,O)$ (\textit{WR}, \textit{RW}, or \textit{WW})\\\hline
\textbf{X}$(O,O)$ & returns true if there is any relation between operation's instances $(O,O)$ (\textit{ST}, \textit{SOT}, \textit{WR}, \textit{RW}, or \textit{WW})\\\hline
\end{tabular}
}
\end{table*}

\subsubsection{Associations between Formula Components}
\label{sec:association_formula_components}

\begin{figure*}[t]
\centering
\scalebox{.8}{\parbox{1.25\linewidth}{%
System Component's Constraints:
\begin{flalign}
    \forall o_1, o_2 \in O &: WR(o_1, o_2)\Rightarrow vis(o_1, o_2) && \label{eq:wr-edge}\\
    \forall o_1, o_2 \in O &: WW(o_1, o_2)\Rightarrow ar(o_1, o_2) && \label{eq:ww-edge} \\
    \forall o_1, o_2 \in O &: RW(o_1, o_2)\Rightarrow \neg vis(o_2, o_1) && \label{eq:rw-edge}\\
    \forall o_1 \in O &: \textit{is\_update}(o_1) \Leftrightarrow(\textit{oname}(o_1) = \textit{Op}_1 \lor \ldots \lor \textit{oname}(o_1) = \textit{Op}_k) && \label{eq:update-oper} \\
    \forall o_1 \in O, &~\textit{Op}_j \in \textit{ONames} : (oname(o_1) = \textit{Op}_j) \Rightarrow ((tname(parent(o_1)) = T(\textit{Op}_j)) && \label{eq:sub-transaction}\\
    \forall o_1 \in O, &~\textit{Op}_j \in \textit{ONames} : (oname(o_1) = \textit{Op}_j) \Rightarrow ((\textit{fname}(origtx(o_1)) = F(\textit{Op}_j)) && \label{eq:original-transaction}\\
    \forall o_1 \in O, &~\textit{Op}_j \in \textit{ONames} : (oname(o_1) = \textit{Op}_j) \Rightarrow (mname(o_1) = M(\textit{Op}_j)) && \label{eq:microservices}\\
    \forall o_1,o_2 \in O &: ar(o_1,o_2) \Rightarrow (otime(o_1) < otime(o_2)) && \label{eq:ar} \\
    \forall o_1,o_2 \in O, &~\textit{Op}_i, \textit{Op}_j \in \textit{ONames},~j > i,~F(\textit{Op}_i) = F(\textit{Op}_j) : \nonumber \\ 
    &~((parent(o_1) = parent(o_2)) \lor (origtx(o_1) = origtx(o_2)) \land (oname(o_1) = \textit{Op}_i)\land (oname(o_2) = \textit{Op}_j)) \Rightarrow (otime(o_1) < otime(o_2)) && \label{eq:sequence}\\
    \forall o_1, o_2 \in O & : D(o_1, o_2)\Rightarrow (\neg (ST(o_1, o_2)\lor SOT(o_1, o_2)) \land (WW(o_1, o_2)\lor WR(o_1, o_2)\lor RW(o_1, o_2))) && \label{eq:operation-dependency}\\
    \forall o_1, o_2 \in O &: X(o_1, o_2) \Rightarrow (ST(o_1, o_2)\lor SOT(o_1, o_2) \lor D(o_1, o_2)) && \label{eq:relations}
\end{flalign}
Consistency Models Constraints:
\begin{flalign}
    \forall o_1, o_2, o_3 \in O & : (ST(o_1, o_2) \land vis(o_1, o_3) \land (mname(o_1) = mname(o_3))) \Rightarrow vis(o_2,o_3) & & \label{eq:read-committed}\\
    \forall o_1, o_2, o_3 \in O & : (ST(o_1, o_2) \land vis(o_3, o_1) \land (mname(o_1) = mname(o_3))) \Rightarrow vis(o_3,o_2) & & \label{eq:repeated_read}\\
    \forall o_1, o_2 \in O & : (ar(o_1, o_2) \land (mname(o_1) = mname(o_2))) \Rightarrow vis(o_1,o_2) & & \label{eq:linearizability}
\end{flalign}
Edge Type Constraints:
\begin{flalign}
    \forall o_1, o_2 \in O & : (parent(o_1) = parent(o_2))\Leftrightarrow ST(o_1, o_2) & & \label{eq:st-edges}\\
    \forall o_1, o_2 \in O & : ((origtx(o_1) = origtx(o_2)) \land (parent(o_1)\neq parent(o_2)))\Leftrightarrow SOT(o_1, o_2) & & \label{eq:sot-edges}
\end{flalign}
Cycle Length Constraints:
\begin{flalign}
   \exists o_1, o_2, \ldots o_k \in O &: Distinct(o_1, o_2, \ldots, o_k) \land (ST(o_1, o_2) \lor SOT(o_1, o_2))\land D(o_2,o_3) \land \ldots \land X(o_i, o_{i+1}) \land  \ldots \land D(o_k,o_1) & & \label{eq:cycles_generic}
\end{flalign}
}}
    \caption{\thesystembold's models constraints} \label{fig:mad-constraints}
\end{figure*}

Let \textit{FNames} = $\{ \textit{Txn}_1, \ldots, \textit{Txn}_f \}$ denote the name set of $f$ original transactions and let \textit{TNames} = $\{ \textit{Txn}_{1,1}, \ldots, \textit{Txn}_{f,t} \}$ denote the name set of sub-transactions where $\textit{Txn}_{i,j}$ refers to the $j^{th}$ sub-transaction in original transaction $\textit{Txn}_i$. Moreover, let \textit{ONames} = $\{\textit{Op}_1, \ldots, \textit{Op}_k\}$ and \textit{MNames} = $\{\textit{M}_1, \ldots, \textit{M}_m\}$ denote the name set of $k$ update operations and $m$ microservices, respectively. For any two operations $\textit{Op}_i$ and $\textit{Op}_j$ from the same transaction where $\textit{Op}_i$ occurs before $\textit{Op}_j$, then we have $j>i$.
Finally, let $F(\textit{Op}_i)$, $T(\textit{Op}_i)$ and $M(\textit{Op}_i)$ denote the names of the original transaction, sub-transaction and microservice of execution for operation $\textit{Op}_i$. Note that all these name sets and name associations are defined through a simple analysis of the program. Afterwards, these are used to encode the relations between the system's components as follows (see Figure~\ref{fig:mad-constraints}):

\medskip\noindent \textbf{C1:} Every two instances of operations related by an \textit{WR} edge have the effects of the first instance visible to the second instance (Equation~\ref{eq:wr-edge}). 

\medskip\noindent \textbf{C2:} Every two instances of operations related by an \textit{WW} edge have the first instance happening before the second instance (Equation~\ref{eq:ww-edge}). 

\medskip\noindent \textbf{C3:} Every two instances of operations related by an \textit{RW} edge have the effects of the second instance not visible to the first instance (Equation~\ref{eq:rw-edge}). 

\medskip\noindent \textbf{C4:} Every instance of an update operation must have an \textit{OName} from the set of update operations names, and vice-versa (Equation~\ref{eq:update-oper}). 

\medskip\noindent \textbf{C5-C7:} Every instance operation with a given \textit{OName} needs to belong to an instance of sub-transaction (specific \textit{TName}), original transaction (specific \textit{FName}) and microservice (specific \textit{MName}) as specified in Equations~\ref{eq:sub-transaction},~\ref{eq:original-transaction} and~\ref{eq:microservices}.

\medskip\noindent \textbf{C8:} Every two instances of operations that have a \textit{ar} relation between them follow an execution order where the first instance occurs before the second instance (Equation~\ref{eq:ar}). 

\medskip\noindent \textbf{C9:} Every two instances of operations that belong to the same original transaction need to follow a sequential execution order according to their order in the Java program (Equation~\ref{eq:sequence}).

\medskip\noindent \textbf{C10:}  Every two instances of operations that have a dependency relation between them (\textit{D}) do not have an \textit{ST} or \textit{SOT} relation and do have an \textit{WW}, \textit{WR} or \textit{RW} relation (Equation~\ref{eq:operation-dependency}). 

\medskip\noindent \textbf{C11:}  Every two instances of operations that have any relation between them (\textit{X}) do have an \textit{ST}, \textit{SOT}, \textit{WW}, \textit{WR} or \textit{RW} relation (Equation~\ref{eq:relations}).

\subsubsection{Consistency Models}
\label{sec:consistency_models}

To make an analysis faithful to the environment where the microservice systems will execute, we assume two consistency models: Serializability and Eventual Consistency. Between operations of the same sub-transaction and other operations of the same microservice, we assume that they will respect Serializability. Between operations of different sub-transactions or that execute on different microservices, we assume Eventual Consistency, as commonly designed in a microservice architecture~\footnote{\url{https://martinfowler.com/articles/microservice-trade-offs.html}}. By default, \thesystem's implementation enforces Eventual Consistency between the visibility effects of the operations. However, since we want to enforce Serializability between operations of the same sub-transaction and other operations of the same microservice, we add assertions to the SMT formula that model the behavior of consistency models. 
The constraints for \textit{Read Committed}, \textit{Repeatable Read} and \textit{Linearizability}~\cite{highly_available_transactions, JepsenConsistencyModels} are presented in Figure~\ref{fig:mad-constraints}, and \textit{Serializability} results from the conjunction of these constraints.

\medskip\noindent \textbf{C12 (Read Committed):} for every three instances of operations $(o_1, o_2, o_3)$, if $o_1$ and $o_2$ belong to the same instance of sub-transaction, the effects of $o_1$ are visible to $o_3$, and $o_1$ and $o_3$ belong to the same microservice, then the effects of $o_2$ are also visible to $o_3$ (Equation~\ref{eq:read-committed}).

\medskip\noindent \textbf{C13: (Repeatable Read):} for every three instances of operations $(o_1, o_2, o_3)$, if $o_1$ and $o_2$ belong to the same instance of sub-transaction, the effects of $o_3$ are visible to $o_1$, and $o_1$ and $o_3$ belong to the same microservice, then the effects of $o_3$ are also visible to $o_2$ (Equation~\ref{eq:repeated_read}).

\medskip\noindent \textbf{C14 (Linearizability):} for every two instances of operations $(o_1, o_2)$, if $o_1$ happens before $o_2$, and $o_1$ and $o_2$ belong to the same microservice, then the effects of $o_1$ are visible to $o_2$ (Equation~\ref{eq:linearizability}).\\

\subsubsection{Types of Edges}

To encode the \textit{ST} and \textit{SOT} edges, \thesystem expresses the following two properties, respectively:

\medskip\noindent \textbf{C15:}  Every two instances of operations that belong to the same instance of sub-trans\-ac\-tion are related via an \textit{ST} edge, and vice-versa (Equation~\ref{eq:st-edges}). 

\medskip\noindent \textbf{C16:}  Every two instances of operations that belong to the same instance of original transaction and do not belong to the same instance of sub-transaction are related via an \textit{SOT} edge, and vice-versa (Equation~\ref{eq:sot-edges}). \\ 

For the dependency edges (\textit{RW}, \textit{WR}, \textit{WW}), constraints are defined for each pair of sub-transactions to establish whether or not it is possible to have a dependency between instances of operations of those sub-transactions. For instance, if two instances of operations $o_1$ and $o_2$ access different tables, then a \textit{WW} can never occur. Otherwise, if both $o_1$ and $o_2$ write in the same table, then a constraint is added to verify if it is possible to have a row where the operations conflict.

\subsubsection{Cycles Assertions}

A cyclic anomalous graph is:

\medskip\noindent \textbf{C17:}  A cycle with at least one \textit{ST} or \textit{SOT} edge and at least two dependency edges (\textit{RW}, \textit{WR}, \textit{WW}). This is captured in Equation~\ref{eq:cycles_generic} where $k$ denotes the size of the cycle.

\begin{figure}[t]
    \centering
    \includegraphics[width=0.23\textwidth]{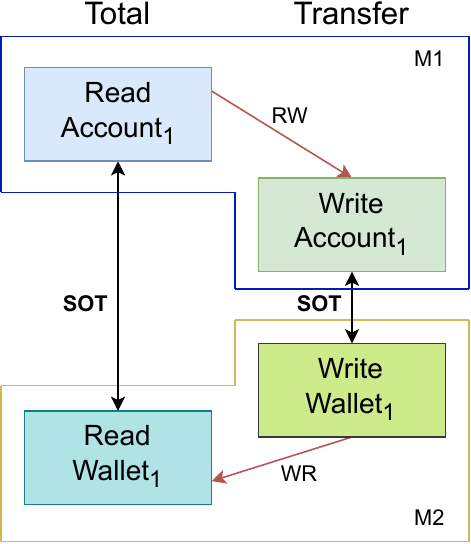}
    \caption{\thesystembold's cyclic graph for the example anomaly.} \label{fig:anomaly_example_graph}
\end{figure}

\medskip Recalling the example anomaly presented in Figure~\ref{fig:anomaly_interleaving}. \thesystem detects that anomaly by finding the cyclic graph that can be seen in Figure~\ref{fig:anomaly_example_graph}. The cycle contains two \textit{SOT} edges and two dependency edges (\textit{RW} and \textit{WR}), and represents the interleaving of the original transactions \textsf{Total} and \textsf{Transfer}.

\subsection{Metrics Extractor}

\thesystem includes a Metrics Extractor component, that gathers information regarding the possible anomalies the microservices application may face when the given decomposition is used. The information collected includes the total number of anomalies, classified in \textit{core anomalies} and \textit{extensions}, the number of anomalies per type and, finally, the entities, functionalities, and sub-transactions involved in each anomaly. We decided to capture these indicators as they allow the programmer to have a better understanding of the potential anomalies and of the amount of effort required to prevent each of them. Furthermore, these figures capture the types of anomalous behaviors to be expected and the combinations that originate those anomalous behaviors.

After the SMT solver finishes its analysis for each subset of functionalities, \thesystem obtains the \textit{Satisfiable Assignments}, which represent the detected anomalies. Using these assignments, \thesystem performs two steps. Firstly, it classifies the obtained assignments into anomaly types, following the definitions of Adya et al~\cite{dependency_graphs}. Secondly, it groups the anomalies by sets of entities, functionalities, and sub-transactions, to understand which combinations can lead to anomalous behaviors. We now dive deeper into the implementation of each step of the \textit{Metrics Extractor}.

To categorize the obtained assignments as anomalies, \thesystem matches the dependency cycle obtained by the assignment with the generalized dependency cycles described in Adya et al~\cite{dependency_graphs}. Furthermore, \thesystem categorizes the anomalies as either \textit{core anomalies} or \textit{extensions}. A \textit{core anomaly} is one whose execution cycle has the minimum size required to express its categorized anomaly. \textit{Extensions} are anomalies whose execution cycle includes a \textit{core anomaly} with additional operations that do not affect the categorized anomaly. This distinction is important since if a developer fixes the source of the \textit{core anomalies}, it will also remove its related \textit{extensions}.

For the second step, \thesystem iterates over the anomalies found and checks the entities, functionalities, and sub-transactions involved for each anomaly. This metric restricts the sources of anomalous behavior to sets of entities/functionalities/sub-transactions, reducing the search space to be explored by the developer to identify the source of the anomalous behavior.

\subsection{Supported Syntax}

In the current version of \thesystem, the source code needs to be a program written in Java using the JDBC syntax. If needed, it is possible to extend the \textit{AR Compiler} to support other additional programming languages. Furthermore, the current version is not able to parse joins or implicit updates. This is not a fundamental limitation, because these operations can be represented through combinations of reads and writes, which are supported by the \textit{AR Compiler} and are enough to cover all benchmarks used in the evaluation. To define a join, one can divide the query with the join operation into two or more queries, each accessing only one table. Similarly, to represent an implicit update, one can divide it into a read, to retrieve the current value, followed by a write with the update expression. These transformations could be implemented by leveraging the current \textit{AR Compiler} or by adding a pre-processing step to automatically decompose the operations. However, this exercise falls outside the scope of this work.

\subsection{Maximum Cycle Length}

As noted before, \thesystem uses the parameter Maximum Cycle Length (MCL) to bound the search space. The default value of MCL is set to $4$, given that the anomalies detected by our tool can be identified by cycles of this length~\cite{ANSI_critique}. By setting MCL to a value that allows detecting all serializability anomalies supported by \thesystem (see Section~\ref{sec:MAD_metrics}), we bound the exploration without generating false negatives.

\section{Evaluation}
\label{chap:evaluation}

\thesystem aims at identifying the anomalies that can emerge when migrating a monolith to microservices following a given decomposition. Our experimental evaluation focuses on the following key research questions:
\textbf{RQ1:} Can \thesystem offer insights regarding the best decompositions?
\textbf{RQ2:} Can \thesystem classify the anomalies to help identifying access patterns that cause the errors?
\textbf{RQ3:} How long  does it take to execute MAD?
\textbf{RQ4:} How effective is the \textit{Divide and Conquer} strategy in improving the \thesystem's performance?

\medskip\noindent We gathered seven benchmarks based on GitHub applications and, using a migration tool~\cite{Mono2Micro_rito}, we have generated two microservices decompositions for each benchmark. By applying \thesystem to these benchmarks and decompositions, we are able to address the research questions above.

\subsection{Experimental Setup}
\label{sec:experimental_benchmarks}

We use benchmarks inspired by applications found on GitHub for the experimental evaluation. Our process to set the benchmarks consists of three steps: 1) gathering monolithic applications from GitHub; 2) adapting them to the syntax processed by \thesystem\ ; 3) generating two microservices decompositions of each application with the help of a migration tool~\cite{Mono2Micro_rito} that supports programmers on the task of grouping the entities by the microservices. The  applications we have used in the evaluation are the following:\\

\medskip\noindent\textbf{TPC-C}\footnote{\url{https://github.com/oltpbenchmark/oltpbench/tree/master/src/com/oltpbenchmark/benchmarks/tpcc}} is defined in the OLTP-Bench~\cite{OLTPBench} project and simulates the behavior of a delivery and warehouse management system;

\medskip\noindent\textbf{FindSportMates}\footnote{\url{https://github.com/chihweil5/FindSportMates}} (findmates) is an application used as a benchmark in other microservices works~\cite{microservices_benchmarks1, microservices_benchmarks3} and simulates a platform where users can manage and find events to connect with other users;

\medskip\noindent\textbf{jpabook}\footnote{\url{https://github.com/holyeye/jpabook/tree/master/ch12-springdata-shop}} simulates a shop where members can order items and track the delivery process;

\medskip\noindent\textbf{JPetStore}\footnote{\url{https://github.com/mybatis/jpetstore-6}} (jpetstore) is an application highly used by previous microservices works~\cite{microservices_benchmarks1, microservices_benchmarks2, Mono2Micro_sac, Mono2micro_microservices_benchmarks4, microservices_benchmarks5} and simulates an online pet store where each user has an account and can browse through a catalog of pets to choose which pets they want to order;

\medskip\noindent\textbf{spring-petclinic}\footnote{\url{https://github.com/spring-projects/spring-petclinic}} (petclinic) is an application used as a benchmark by a previous microservices work~\cite{microservices_benchmarks2} and simulates the operation of a pet clinic;

\medskip\noindent\textbf{myweb}\footnote{\url{https://github.com/Jdoing/myweb}} is an application that simulates the behavior of the web allowing users to have roles and manage resources;

\medskip\noindent\textbf{spring-mvc-react}\footnote{\url{https://github.com/noveogroup-amorgunov/spring-mvc-react}} (react) is a platform where users can post questions and answers with tags associated with them. Besides that, the system also allows users to upvote or downvote publications, which influences the users' popularity.

\medskip We have selected these benchmarks because they cover a wide range of domain areas and have implementations that address real-world scenarios and, yet, they are simple enough to be processed by the \thesystem prototype in a reasonable time. 

For each benchmark, we analyze three decompositions. First, the \textit{mono} decomposition, which represents the monolithic version of the benchmark and shows the initial number of anomalies. For this decomposition, the number of sub-transactions will always be the same as the number of functionalities since there is no division of transactions in the monolithic version. We assume that the monolith is correct and contains no anomalies, so any anomaly arising in the microservices' decompositions must have resulted from the migration. Second, the \textit{``best''} decomposition, which is the one with the highest \textit{Silhouette Score} (a metric used to assess how well the clustering of the entities is done) calculated by the migration tool~\cite{Mono2Micro_rito}. At last, in the \textit{full} decomposition, each entity is managed by a different microservice, which is the worst-case scenario in terms of potential anomalies.

In this context, we use \thesystem considering four as the maximum cycle length. The process to choose this value consisted of starting with value three since it is the minimum number of edges required to detect a cycle with an anomaly, and incrementing it to ensure that all core anomaly types could be detected (requires length four~\cite{ANSI_critique, CLOTHO}), while still assuring that \thesystem was able to analyze the benchmarks within a timeout limit of 4 hours (14400 seconds). We defined this timeout limit as the reasonable amount of time a programmer would wait for the analysis to be complete. The evaluation was performed in a virtual machine with 32 virtual CPU cores running on two Intel(R) Xeon(R) Gold 5320 CPUs at 2.2GHz and 128GB of DDR4 RAM with Intel Optane Memory configured in App Mode. The virtual machine uses Ubuntu 18.04.4 LTS, Java 8, and version 4.12.3 of Z3 with the default configuration. 


\begin{table}[t]
\centering
\caption{Anomalies detected.}\label{tab:overall_results}
\scalebox{0.6}{
\begin{tabular}{|l r r c r r||r r r|}
\hline
\bfseries  & & & & & & \multicolumn{3}{c|}{\bfseries MAD} \\
\bfseries Benchmark & \bfseries \#E & \bfseries \#F & \bfseries Config & \bfseries \#M & \bfseries \#ST & \multicolumn{1}{c|}{\bfseries \#CA} & \multicolumn{1}{c|}{\bfseries \#TA} & \multicolumn{1}{c|}{\bfseries ET (s)}\\
\hline
\hline
& & & mono & 1 & 5 &  0 & 0 & 10\\

\bfseries TPC-C & 9 & 5 & ``best'' & 2 & 6 &  0 & 0 & 9\\

& & & full & 9 & 22 & 28 & 98 & 306\\
\hline
\hline
& & & mono & 1 & 9 &  0 & 0 & 23\\

\bfseries findmates & 2 & 9 & ``best''/ full& 2 & 13 &  0 & 0 & 31\\
& & & & & & & &\\ 
\hline
\hline
& & & mono & 1 & 10 &  0 & 0 & 48\\

\bfseries jpabook & 5 & 10 & ``best'' & 2 & 15 &  22 & 70 & 136\\

& & & full & 5 & 21 &  25 & 79 & 222\\
\hline
\hline
& & & mono & 1 & 11  & 0 & 0 & 454\\

\bfseries jpetstore & 13 & 11 & ``best'' & 3 & 17  & 40 & 52 & 7\\

& & & full & 13 & 48 &  85 & 188 & 10\\
\hline
\hline
& & & mono & 1 & 12 & 0 & 0 & 93\\

\bfseries petclinic & 6 & 12 & ``best'' & 2 & 14 &  0 & 0 & 89\\

& & & full & 6 & 27  & 0 & 0 & 128\\
\hline
\hline
& & & mono & 1 & 20 &  0 & 0 & 309\\

\bfseries myweb & 5 & 20 & ``best'' & 2 & 22 &  0 & 0 & 358\\

& & & full & 5 & 32 & 7 & 7 & 514\\
\hline
\hline
& & & mono & 1 & 23 &  0 & 0 & 545\\

\bfseries react & 5 & 23 & ``best'' & 2 & 28 &  13 & 45 & 996\\

& & & full & 5 & 39 &  18 & 58 & 2\\
\hline
\multicolumn{9}{c}{(E=entities; F=functionalities; M=microservices; ST=sub-transactions; )}\\
\multicolumn{9}{c}{CA=core anomalies; TA=total anomalies; ET=execution time)}
\end{tabular}
}
\end{table}

\subsection{RQ1: Insights Gained from Using MAD}
\label{sec:insights}

Table~\ref{tab:overall_results} presents the results obtained when applying \thesystem to our benchmarks. By looking at the number of anomalies found in each decomposition, \thesystem allows the programmers to assess how problematic each decomposition will be, therefore enabling them to make a more informed decision when migrating to microservices. For instance, in \textit{jpabook}, \textit{jpetstore}, and \textit{react}, even their \textit{``best''} decompositions have anomalies. This occurs because the \textit{``best''} decompositions for these cases require the functionalities to be chopped into multiple sub-transactions, opening the door for more interleaving between the functionalities. Note also that all microservices decompositions, except \textit{jpetstore} \textit{best} and \textit{myweb} \textit{full}, have a fairly higher number of anomalies when compared with the number of \textit{core anomalies}. Thus, a fair portion of the anomalies found (between 54 and 72 percent) are simply \textit{extensions} of smaller number of core anomalies and, therefore, can be eliminated if the core anomalies are eliminated. The \textit{findmates} and \textit{petclinic} benchmarks have no anomalies in any of their decompositions because their operations are mostly reads and the functionalities tend to be short. This, respectively, leads to fewer conflicts between accesses and fewer interleavings between functionalities.

Another aspect one can notice in the table is that all the \textit{``best''} decompositions, except for \textit{jpetstore}, have two microservices. From our understanding, this phenomenon is possibly related to Martin Fowler's Strangler Fig pattern,\footnote{\url{https://martinfowler.com/bliki/StranglerFigApplication.html}} where the migration to microservices is done incrementally by extracting a few services at a time, considering the coupling between entities. The \textit{Silhouette Score} might indirectly take this into account, resulting in its value suggesting that from the monolithic implementation, the most appropriate decomposition to microservices is to migrate to two microservices. 

\subsection{RQ2: Classifying Anomalies}
\label{sec:MAD_metrics}

We now show the results for the anomaly classification task performed by \thesystem. In particular, we show the number of anomalies classified by type and by set of sub-transactions.
Being aware of the anomaly types involved in a decomposition allows developers to gain better insight on the possible costs and work required to mitigate the effects of anomalies in a given decomposition.
For example, a decomposition with only \textit{Non-Repeatable Reads} could be mitigated with a light-weight weakly consistent transactional protocol like Transactional Causal Consistency~\cite{akkoorath2016cure} while \textit{Lost Updates} or \textit{Write Skews} require a strongly consistent transactional protocol like Snapshot Isolation or Strict Serializability~\cite{JepsenConsistencyModels}.

Table~\ref{tab:results_per_type} shows the number of anomalies found for each decomposition by type of anomaly.  We are only considering the analysis of the sub-transactions metrics for one decomposition, but this can be done for all decompositions. Note that we omit the results for the \textit{findmates} and \textit{petclinic} benchmarks since they have no anomalies in any decomposition. Also, the \textit{Write Skews} are counted together with the \textit{Lost Updates} because the write skew pattern also corresponds to one of the lost update patterns. The only way to distinguish between them would be to also consider the rows accessed. If the same row was being accessed, then it would be a \textit{Lost Update}. Otherwise, it would be a \textit{Write Skew}. However, the current patterns used in our approach only consider the graph cycle edges and operation types, and does not include information regarding the accessed rows which is required to make this distinction. 

\begin{table}[t]
\centering
\caption{\thesystem anomalies found per type.}\label{tab:results_per_type}
\scalebox{0.6}{

\begin{tabular}{|l c r r r r r r r r||r|}
\hline
\bfseries Benchmark & \bfseries Config & \multicolumn{1}{c}{\bfseries \#DR} & \multicolumn{1}{c}{\bfseries \#DW} & \multicolumn{1}{c}{\bfseries \#LU} & \multicolumn{1}{c}{\bfseries \#LU/} & \multicolumn{1}{c}{\bfseries \#NRR} & \multicolumn{1}{c}{\bfseries \#PR} & \multicolumn{1}{c}{\bfseries \#RS} & \multicolumn{1}{c||}{\bfseries \#Ext} & \multicolumn{1}{c|}{\bfseries \#Total} \\
 & & \multicolumn{1}{c}{\bfseries } & \multicolumn{1}{c}{\bfseries } & \multicolumn{1}{c}{\bfseries } & \multicolumn{1}{c}{\bfseries WS} & \multicolumn{1}{c}{\bfseries } & \multicolumn{1}{c}{\bfseries } & \multicolumn{1}{c}{\bfseries } & &\\
\hline
\hline
& mono &  &  &  &  &  &  &  &  & 0 \\

\bfseries TPC-C & ``best'' &  &  &  &  &  &  &  &  & 0 \\

& full &  & 13 &  & 3 & & & 12 & 70 & 98 \\
\hline
\hline
& mono &  &  &  &  &  &  &  &  & 0 \\

\bfseries jpabook & ``best'' &  &  & 3 & 10 & & & 9 & 48 & 70 \\

& full &  &  & 3 & 11 & & & 11 & 54 & 79 \\
\hline
\hline
& mono &  &  &  &  &  &  &  &  & 0 \\

\bfseries jpetstore & ``best'' &  & 4 &  & 12 &  &  & 24 & 12 & 52\\

& full &  & 7 &  & 14 & 2 &  & 62 & 103 & 188 \\
\hline
\hline
& mono &  &  &  &  &  & & & & 0 \\

\bfseries myweb & ``best'' &  & & & & &  &  &  & 0 \\

& full &  &  &  & 3 & & & 4 & & 7 \\
\hline
\hline
& mono &  &  &  &  &  & & & & 0 \\

\bfseries react & ``best'' &  &  &  &  & 1 & 1 & 11 & 32 & 45 \\

& full &  &  &  &  & 3 & 4 & 11 & 40 & 58 \\
\hline
\multicolumn{11}{c}{(DR=dirty read, DW=dirty write; LU=lost update; WS=write skew, }\\
\multicolumn{11}{c}{NRR=non-repeatable read; PR=phantom read; RS=read skew, Ext=extensions)}
\end{tabular}
}
\end{table}

In Table~\ref{tab:tpcc_full_core_anomalies_entities}, we present the \textit{core anomalies} found in the \textit{TPC-C} \textit{full} decomposition. Developers may leverage this information to guide their decomposition design, identifying the most costly entity decouplings. For instance, notice that the combination of entities \textit{[oorder, order\_line]} are heavily coupled, as decomposing these entities in different microservices generates five new anomalies. Merging these entities in the same microservice could significantly reduce anomaly mitigation costs when migrating the application.

Furthermore, in Table~\ref{tab:tpcc_full_core_anomalies_sub_transactions}, we present the combinations of sub-transactions that originate the \textit{core anomalies} in the decomposition. This information highlights the key sections in the application that require more care when migrating the application, allowing developers to be better informed during the decomposition process on what anomalies they will face and what kind of techniques will be required to mitigate the effects of said anomalies.

\begin{table}[t]
\centering
\caption{\textit{TPC-C} \textit{full} \textit{core anomalies} entities.}\label{tab:tpcc_full_core_anomalies_entities}
\scalebox{0.6}{
\begin{tabular}{|c r c|}
\hline
\bfseries Entities & \bfseries \#Anomalies & \bfseries Anomalies Types\\
\hline
\hline
[oorder, order\_line] & 5 & [DW, RS]\\

[customer, district] & 4 & [DW]\\

[customer, warehouse] & 4 & [DW]\\

[new\_order, order\_line] & 4 & [LU/WS, RS]\\

[new\_order, oorder] & 3 & [LU/WS, RS]\\

[customer, new\_order] & 2 & [LU/WS, RS]\\

[customer, oorder] & 1 & [DW]\\

[customer, order\_line] & 1 & [DW]\\

[district, order\_line] & 1 & [RS]\\

[district, stock] & 1 & [DW]\\

[district, warehouse] & 1 & [DW]\\

[order\_line, stock] & 1 & [RS]\\
\hline
\multicolumn{3}{c}{(DW=dirty write; LU=lost update; WS=write skew, RS=read skew}
\end{tabular}
}
\end{table}

\begin{table}[t]
\centering
\caption{\textit{TPC-C} \textit{full} \textit{core anomalies} sub-transactions.}\label{tab:tpcc_full_core_anomalies_sub_transactions}
\scalebox{0.49}{
\begin{tabular}{|c c c c |}
\hline
\bfseries Functionalities & \bfseries Sub-transactions & \bfseries \#Anomalies & \bfseries Anomalies Types\\
\hline
\hline
[payment] & [payment\_0, payment\_2] & 4 & [DW]\\

[payment] & [payment\_1, payment\_2] & 4 & [DW]\\

[delivery] & [delivery\_0, delivery\_1] & 2 & [LU/WS, RS]\\

[delivery] & [delivery\_0, delivery\_2] & 2 & [LU/WS, RS]\\

[delivery] & [delivery\_0, delivery\_3] & 2 & [LU/WS, RS]\\

[orderStatus, newOrder] & [orderStatus\_1, orderStatus\_2, newOrder\_3, newOrder\_7] & 2 & [RS]\\

[delivery, newOrder] & [delivery\_0, delivery\_2, newOrder\_4, newOrder\_7] & 2 & [RS]\\

[delivery, newOrder] & [delivery\_1, delivery\_2, newOrder\_3, newOrder\_7] & 2 & [RS]\\

[payment] & [payment\_0, payment\_1] & 1 & [DW]\\

[newOrder] & [newOrder\_2, newOrder\_6] & 1 & [DW]\\

[delivery] & [delivery\_1, delivery\_2] & 1 & [DW]\\

[delivery] & [delivery\_1, delivery\_3] & 1 & [DW]\\

[delivery] & [delivery\_2, delivery\_3] & 1 & [DW]\\

[newOrder, stockLevel] & [stockLevel\_0, stockLevel\_1, newOrder\_2, newOrder\_7] & 1 & [RS]\\

[newOrder, stockLevel] & [stockLevel\_1, stockLevel\_2, newOrder\_6, newOrder\_7] & 1 & [RS]\\

[delivery, newOrder] & [delivery\_0, delivery\_1, newOrder\_3, newOrder\_4] & 1 & [RS]\\
\hline
\multicolumn{4}{c}{(DW=dirty write; LU=lost update; WS=write skew, RS=read skew}
\end{tabular}
}
\end{table}

\subsection{RQ3: MAD Execution Time}
\label{sec:execution time}

The last column of  Table~\ref{tab:overall_results} depicts the time required to execute MAD on each decomposition. The execution time of \textit{CMMAM} is not presented in the table, since it is negligible ($\approx$0s). As can be observed, although \thesystem can analyze these cases with more precision than \textit{CMMAM}, \thesystem often requires non-negligible time to perform the analysis. \thesystem does not require long executions for simple applications nor when the decompositions do not use many sub-transactions. However, for complex applications with a high number of functionalities and/or sub-transactions, \thesystem needs to analyze a large number of combinations of transactions, therefore taking more time to finish its execution (in our experiments, the longest execution time was observed when analyzing the full decomposition of the \textit{jpetstore}, that took almost 3 hours).

\subsection{RQ5: Divide and Conquer Performance}
\label{sec:search_algorithm_impact}

To evaluate if \thesystem's divide and conquer strategy improves the performance of \thesystem's analysis and makes the analysis of more complex applications/decompositions feasible, we compare \thesystem's performance with and without the divide and conquer strategy when applied to the same decompositions. To assess the impact of the parallelization, we also consider the performance of the divide and conquer strategy single-threaded and multi-threaded.

\begin{table}[t]
\centering
\caption{Divide and conquer performance.}\label{tab:search_algorithm_impact}
\scalebox{0.7}{
\begin{tabular}{|l c r r r|}
\hline
\bfseries Benchmark & \bfseries Decomposition & \multicolumn{1}{c}{\bfseries NO-DC [s]} & \multicolumn{1}{c}{\bfseries S-DC [s]} & \multicolumn{1}{c|}{\bfseries P-DC [s]}\\
\hline
\hline
& mono & 7 & 18 & 10\\

\bfseries TPC-C & ``best'' & 7 & 20 & 9\\

& full & 1,715 & 339 & 306\\
\hline
\hline

& mono & 4 & 20 & 23\\

\bfseries findmates & ``best''/full & 5 & 24 & 31\\
&  &  &  & \\
\hline
\hline

& mono & 7 & 37 & 48\\

\bfseries jpabook & ``best'' & 4,263 & 220 & 136\\

& full & 5,342 & 286 & 222\\
\hline
\hline

& mono & 38 & 850 & 454\\

\bfseries jpetstore & ``best'' & (timeout) & 7,670  & 7,239\\

& full & (timeout) & 10,435 & 10,249\\
\hline
\hline

& mono & 10 & 67 & 93\\

\bfseries petclinic & ``best'' & 11 & 68 & 89\\

& full & 13 & 92 & 128\\
\hline
\hline

& mono & 11 & 218 & 309\\

\bfseries myweb & ``best'' & 11 & 227 & 358\\

& full & 213 & 353 & 514\\
\hline
\hline

& mono & 153 & 406 & 545\\

\bfseries react & ``best'' & (timeout) & 982 & 996\\

& full & (timeout) & 1,209 & 1,517\\
\hline
\multicolumn{5}{c}{NO-DC: without divide and conquer; }\\
\multicolumn{5}{c}{S-DC: sequential divide and conquer; P-DC: parallel divide and conquer}
\end{tabular}
}
\end{table}

The results in Table~\ref{tab:search_algorithm_impact} show that the divide and conquer strategy can significantly shorten the analysis time. Gains are more significant for complex cases since each SMT formula is much smaller, and it mitigates the time and space complexities by not considering all the original transactions simultaneously. As a result, all analysis can be performed within the time limit (4 hours = 14,400 seconds). Without this strategy, \thesystem exceeds the timeout when analyzing decompositions \textit{``best''} and \textit{full} of the \textit{jpetstore} and \textit{react} benchmarks. However, for simple cases, \thesystem's performance tends to be worse with the divide and conquer strategy. This occurs because the strategy originates an overhead to \thesystem's analysis by requiring unnecessary iterations over combinations with no anomalies. In those simple cases, \thesystem would analyze the interleavings between all the original transactions faster without the strategy because it could consider all of them simultaneously without facing a high number of possible combinations. We also note that the strategy is not fully parallelizable, since the threads of bigger size combinations need to wait for all the threads of smaller size combinations to finish in order to start. 
Therefore, the analysis time for a given combination size is bounded by the analysis time of the slowest thread that is analyzing a combination of that size.

\section{Conclusions and Future Work}

This paper has presented \thesystem, the first framework that automatically detects anomalies that result from the migration of a monolith application to a microservices architecture. \thesystem avoids the under/overestimation limitations of previous works and can classify the anomalies according to the access patterns that cause them, thus helping the programmer by identifying scenarios that result in anomalies. Experimental results from applying \thesystem to different decompositions of benchmarks inspired by applications on GitHub show that \thesystem can offer insights on the complexity of a decomposition that are  more precise than the metrics extracted by related work.  We plan to extend the framework in several ways. For instance, we will add support for associations between entities in different microservices, such as JPA relationships, foreign keys, and semantic invariants that the system must respect. Moreover, we will ease the comparison between different decompositions of the monolith and leverage the current architecture to optimize anomaly detection in a set of decompositions to microservices.

\subsubsection*{Acknowledgements}

This research partially supported by  the Fundação para a Ciência e a Tecnologia (FCT) via scholarship UI/BD/153590/2022,  the INESC-ID grant UIDB/50021/2020 (DOI:10.54499/UIDB/50021/2020) and the DACOMICO project (financed by the OE with ref. PTDC/CCI-COM-
/2156/2021).

\end{document}